\documentclass[aps, twocolumn,noshowpacs,preprintnumbers,amsmath,amssymb]{revtex4}

\usepackage{verbatim}
\usepackage{wasysym}

\usepackage{graphicx}
\usepackage{dcolumn}
\usepackage{bm}
\usepackage{wrapfig}

\begin{document}

\title{Adhesion and size dependent friction anisotropy in boron nitride nanotubes}

\author{Hsiang-Chih Chiu}
\affiliation{School of physics, Georgia Institute of Technology, Atlanta, USA}
\author{Sedat Dogan}
\author{Mirjam Volkmann}
\author{Christian Klinke}
\affiliation{Institute of Physical Chemistry, University of Hamburg, Hamburg, Germany}
\author{Elisa Riedo}
\affiliation{School of physics, Georgia Institute of Technology, Atlanta, USA }

\begin{abstract} 

The frictional properties of individual multiwalled boron nitride nanotubes (BN-NTs) synthesized by chemical vapor deposition (CVD) and deposited on a silicon substrate are investigated using an atomic force microscope tip sliding along (longitudinal sliding) and across (transverse sliding) the tube's principal axis. Because of the tube transverse deformations during the tip sliding, a larger friction coefficient is found for the transverse sliding as compared to the longitudinal sliding. Here, we show that the friction anisotropy in BN-NTs, defined as the ratio between transverse and longitudinal friction forces per unit area, increases with the nanotube-substrate contact area, estimated to be proportional to $(L_{NT} \cdot R_{NT})^{1/2}$, where $L_{NT}$ and $R_{NT}$ are the length and the radius of the nanotube, respectively. Larger contact area denotes stronger surface adhesion, resulting in a longitudinal friction coefficient closer to the value expected in absence of transverse deformations. Compared to carbon nanotubes (C-NT), BN-NTs display a friction coefficient in each sliding direction with intermediate values between CVD and arc discharge C-NTs. CVD BN-NTs with improved tribological properties and higher oxidation temperature might be a better candidate than CVD C-NTs for applications in extreme environments.

\end{abstract}

\maketitle

\section*{Introduction}

\begin{figure}[ht]
  \centering
  \includegraphics[width=0.4\textwidth]{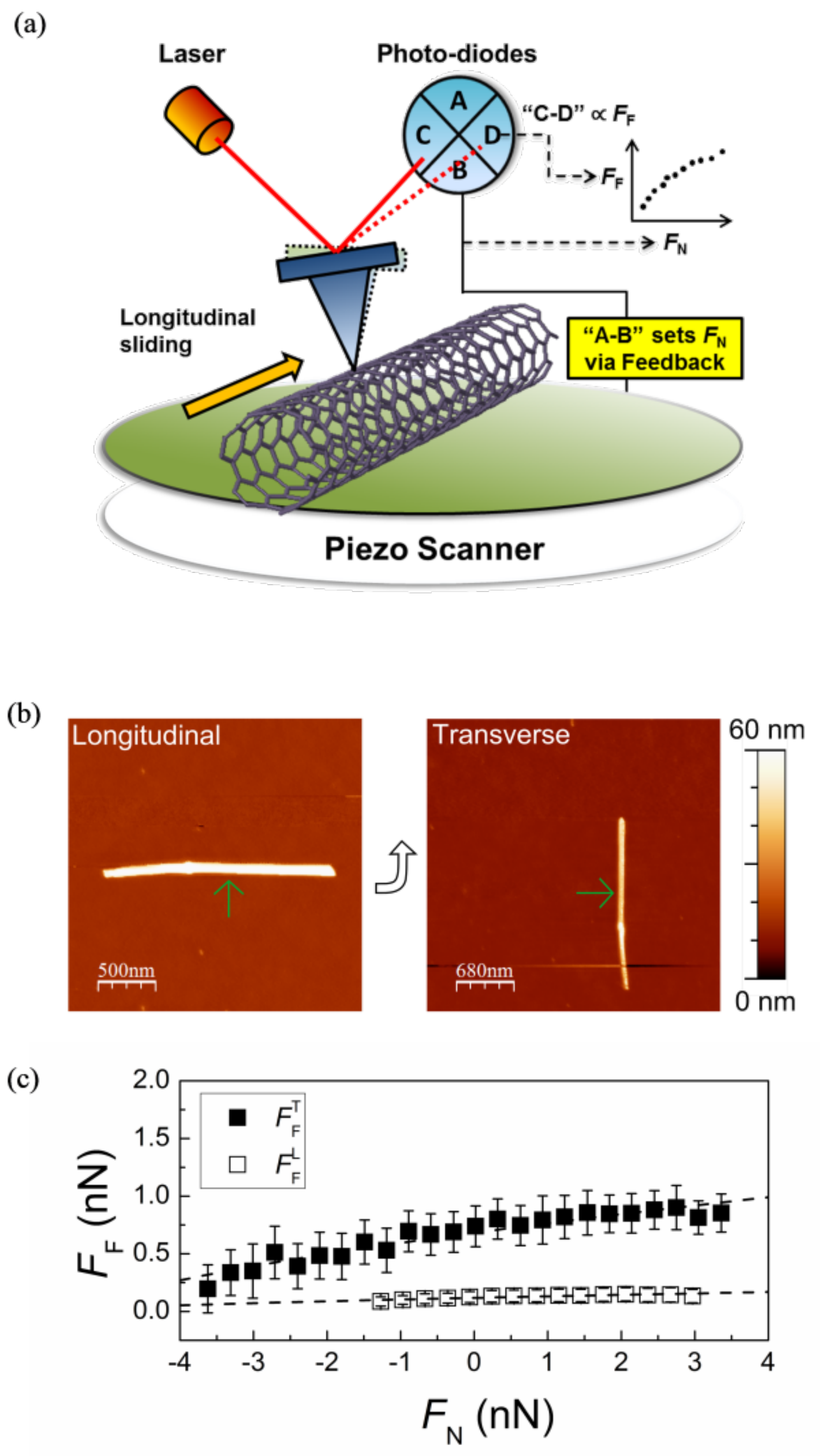}
  \caption{\textit{(a) Experimental setup for the friction force measurement on a nanotube with an AFM during longitudinal tip sliding. The vertical deflection signal of the cantilever determines the normal load $F_{N}$, while the horizontal deflections signal will be proportional to the friction force, $F_{F}$, with which the cantilever encounters during sliding; (b) AFM images of a typical MW BN-NT on a silicon substrate. The green arrows indicate the portion of nanotube where the friction measurements are performed; (b) Friction force along the longitudinal and transverse directions as a function of normal load FN for the BN-NT shown in Figure 1(c). This BN-NT has a radius of $13.9 \pm 0.1$ nm (determined from half of the height of BN-NT's cross sectional profile in the image) and a friction anisotropy of $5$.}}
\end{figure}

Understanding the frictional behaviour of novel nanomaterials such as nanotubes is crucial to the development of nanoscale devices such as nano-electro-mechanical systems (NEMS) or nanocomposites. Nanomaterials such as carbon nanotubes (C-NTs) have been extensively studied in the last decade while its sister counterpart, boron nitride nanotubes (BN-NTs), are recently attracting more and more attention due to their exceptional physical properties. Similar to C-NTs, BN-NTs have a layered structure with alternating boron and nitride atoms in a honeycomb configuration. Since their first successful synthesis in 1995 \cite{1}, significant research efforts have been devoted to understand their extraordinary physical properties. BN-NTs are the strongest insulators ever discovered with a wide band gap $\approx$ 5.5 eV. Their tensile Young's moduli are reported to be between 0.5 and 1.2 TPa by several groups using Transmission Electron Microscope (TEM) \cite{1,2,3,4,5}. Furthermore, BN-NTs are chemically inert and have a high thermal conductivity with an oxidation temperature of 800$^{\circ}$C, compared to only 400$^{\circ}$C for C-NTs \cite{6}. Moreover, the broken sublattice symmetry of BN-NTs along the axis was theoretically shown to produce macroscopic electrical polarization \cite{7}. Because of this polarization, BN-NTs have a stronger interaction with a polymer matrix as experimentally demonstrated in the polyaniline (PANI) composites when compared with C-NTs \cite{8}. It has also been demonstrated that a small addition of BN-NTs in engineered ceramics can drastically enhance their superplastic deformation at high temperatures \cite{9}. Additionally, BN-NTs have been proposed as reinforcement in bio-degradable materials for orthopedic applications \cite{10}. Despite the great potential applications, the tribological properties of BN-NTs have never been reported in literature, differently from multiwalled (MW) C-NTs, whose frictional behaviour has been extensively studied both theoretically \cite{11} and experimentally by Atomic Force Microscopy (AFM) \cite{12,13,14,15,16} and TEM \cite{17,18}. Recent experimental investigations based on AFM have shown that structural defects, chirality and even surface functionalization of C-NTs can significantly influence their frictional properties \cite{12,13}. More specifically, a nano-size silicon AFM tip has been used to slide along (longitudinal sliding, L) and across (transverse sliding, T) a nanotube lying on a silicon substrate. A larger friction coefficient was found during the transverse sliding as compared with the longitudinal one. This friction anisotropy was explained by a transversal deformation or "hindered rolling" of the nanotube, which opens an additional friction dissipation channel occurring mainly during the transverse sliding, leading to a higher friction force. Such dissipation mechanism is partially absent during the longitudinal tip sliding, In addition, the friction anisotropy, defined by the ratio of the transverse and longitudinal friction forces per unit area, is found to be strongly correlated with the presence of structural defects, chirality and surface functionalization. Here, we report on the measurements of the frictional properties of individual MW BN-NTs synthesized by chemical vapor deposition (CVD). A silicon AFM tip is used to slide on top of a BN-NT deposited on a silicon substrate to study the sliding direction dependence of the friction forces. We find that the friction anisotropy in BN-NTs increases quasi-linearly with the nanotube-substrate contact area, estimated by using the Hertz model to be proportional to $(L_{NT} \cdot R_{NT})^{1/2}$, where $L_{NT}$ and $R_{NT}$ are the length and the radius of the nanotube, respectively. A larger contact area denotes a stronger surface adhesion, which decouples the longitudinal sliding with the transversal deformations, resulting in lower longitudinal shear strength. Furthermore, we compare these results with similar measurements previously performed on MW C-NTs grown by arc discharge (AD) and CVD methods \cite{12}. We find that the measured shear strengths of CVD BN-NTs have intermediate values between those of AD C-NTs and CVD C-NTs, probably owing to different amount of structural defects present in the nanotubes. The maximum friction anisotropy for CVD BN-NTs is found to be equal to 8.4, smaller than the value found for AD CNTs, probably due to the absence of armchair BN-NTs. A simple analytical model \cite{12} is applied to the measured friction data to calculate the transverse-longitudinal coupling parameter $\alpha$, the intrinsic shear strength $\sigma^{int}$, which can be understood as the shear strength between a tip and a $h$-BN film, and the hindered rolling shear strength $\sigma^{HR}$, which describes the additional dissipation due to the transversal deformations. Finally, we compare all these key parameters describing the frictional behaviour of BN-NTs with both AD and CVD C-NTs.

\begin{figure}[ht!]
  \centering
  \includegraphics[width=0.4\textwidth]{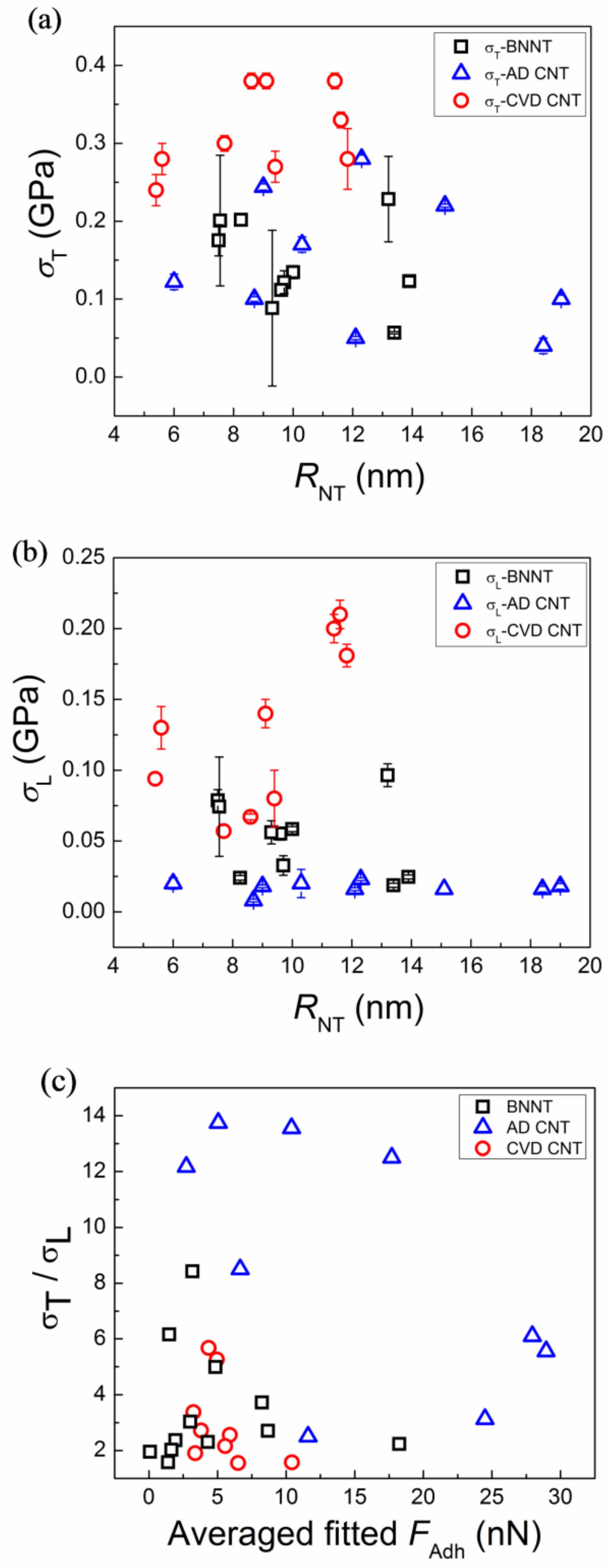}
  \caption{\textit{(a) Transverse shear strengths $\sigma_{T}$ of all the studied BN-NTs and C-NTs; (b) Longitudinal shear strengths $\sigma_{L}$ of all the studied BN-NTs and C-NTs. (c) The friction anisotropy $\sigma_{T} / \sigma_{L}$ vs. the fitted adhesion force $F_{Adh}$ by using equation (1).}}
\end{figure}

\section*{Experimental Methods}

Figure 1(a) shows the schematic of the experimental setup for typical friction force measurements with an AFM. The friction force is measured by the torsional motion of the AFM cantilever when it is sliding on top of a nanotube. This torsional motion is monitored by a laser beam reflected from the back of the cantilever and recorded by a 4 quadrant photo-diodes as shown in the figure, while the constant normal loads during friction measurements are controlled by the feedback loop of the AFM system through the vertical deflection signals. The CVD MW BN-NTs used in this work are purchased from Nanotechlabs, Inc. TEM and AFM images (figure 1 (b)) show that these MW BN-NTs are mostly straight and have few structural defects, similarly to AD C-NTs. For the AFM characterization, these BN-NTs are deposited on an untreated silicon substrate following a procedure similar to the ones reported in Ref. \cite{12,13}. The topography and friction measurements on individual BN-NTs are obtained simultaneously by means of a Veeco Multimode AFM. A nano-size, silicon AFM tip (Nano and More, ppp-LFMR) is used to slide on top of a BN-NT in both the transverse and longitudinal directions for the acquisition of friction forces, $F_{F}$, and the T-L friction anisotropy, $a$, which is defined as the ratio between the shear strengths in the two sliding directions. The spring constant $k_{N} \approx 0.2$ N/m is calibrated by the Sadar's method \cite{19}. The lateral sensitivity of the cantilever is calibrated by the Wedge method \cite{20}. The tip radius $R_{tip} \approx 50$ nm and nanotube radius $R_{NT}$ are inferred directly from the AFM images of nanotube in tapping mode \cite{21}. We first operate the cantilever in tapping mode to find a BN-NT. After an appropriate BN-NT is located, the substrate is first rotated until the axis of the BN-NT is parallel to the fast scan direction of the AFM cantilever. In this arrangement, we zoom into the desired section of the nanotube and measure the longitudinal friction forces in contact mode. In the following step, the sample is rotated by 90$^{\circ}$ for the transverse friction measurements on the same portion of the BN-NT to ensure similar contributions to the friction forces from defects as well as its chirality. The tip velocity is always kept at 1 $\mu$m/s for all measurements. This process is illustrated in figure 1(b) with the AFM images of a typical BN-NT. All measurements are performed with a relative humidity around 40\%.

\begin{figure}[ht]
  \centering
  \includegraphics[width=0.4\textwidth]{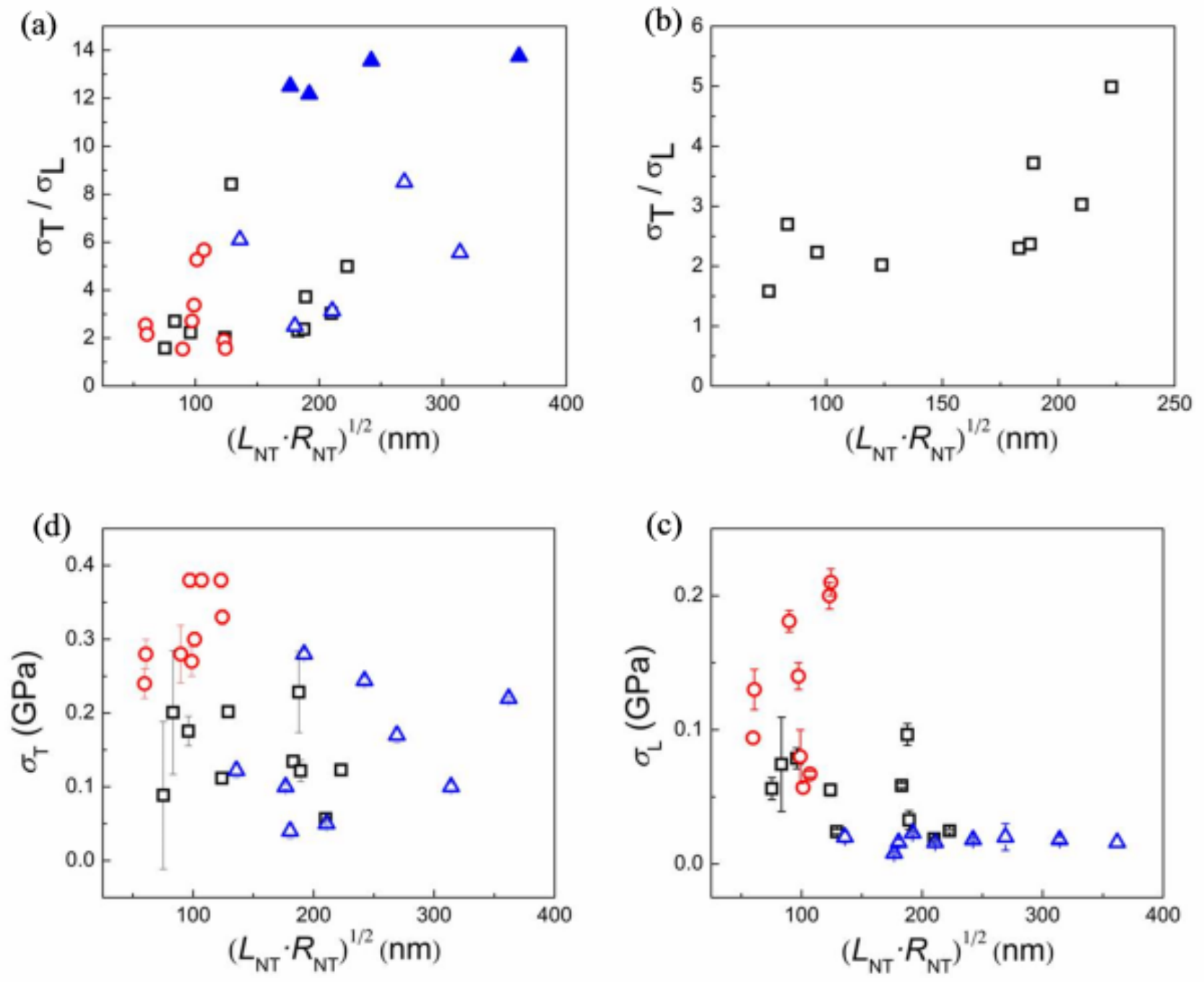}
  \caption{\textit{(a) Friction anisotropy, $\sigma_{T} / \sigma_{L}$ , of all studied BN-NTs (square), AD C-NTs (triangle) and CVD C-NTs (circle) as a function of $( L_{NT} \cdot R_{NT} )^{1/2}$. The solid triangles represent AD C-NTs with very large friction anisotropy ($a > 12$) which might be non-chiral or generally more symmetric nanotubes; (b) A zoom-in of Figure 3(a) for only BN-NTs; (c) and (d) The transverse and longitudinal shear strengths $\sigma_{T,L}$ of all studied nanotubes as function of $( L_{NT} \cdot R_{NT} )^{1/2}$.}}
\end{figure}

\section*{Results and discussion}

Figure 1(c) shows the friction force as a function of the normal load, $F_{N}$, measured on a BN-NT during the longitudinal and transverse sliding, respectively. It is clear that the transverse friction force is larger than the longitudinal one, mainly due to the transversal deformations, i.e. hindered rolling, of the nanotubes during the transverse sliding. In this figure, the transverse friction $F^{T}_{F}$ is approximately 5 times larger than the longitudinal one $F^{L}_{F}$. 

\begin{table*}
\caption{The measured maximum and minimum friction anisotropy ($a$), average transverse and longitudinal shear strengths ($\sigma_{T}$, $\sigma_{L}$ (GPa)), coupling parameter ($\alpha$), intrinsic shear strength ($\sigma^{int}$ (GPa)) and "hindered rolling" shear strengths ($\sigma^{HR}$ (GPa)) of BN-NTs as well as the C-NTs from Ref. \cite{12}.}
    \begin{tabular}{|c|c|c|c|c|c|c|c|} 
				\hline
        \textbf{Sample}    & $a^{Max}$& $a^{Min}$ & $< \sigma_{T} >$ & $< \sigma_{L} >$ & $< \sigma^{int} >$ & $< \alpha >$ & $< \sigma^{HR} >$ \\ \hline
        \textbf{AD C-NT}   & 13.7	& 2.5	& 0.15 $\pm$ 0.03	& 0.02 $\pm$ 0.001	& 0.005 						& 0.10 $\pm$ 0.02 & 0.15 $\pm$ 0.03 \\ \hline
        \textbf{CVD BN-NT} & 8.4	& 1.6	& 0.21 $\pm$ 0.05	& 0.08 $\pm$ 0.02		& 0.007 $\pm$ 0.002	& 0.25 $\pm$ 0.03	& 0.18 $\pm$ 0.02 \\ \hline
        \textbf{CVD C-NT}  & 5.7	& 1.5	& 0.31 $\pm$ 0.02	& 0.13 $\pm$ 0.02		& 0.022 $\pm$ 0.003	& 0.25 $\pm$ 0.03	& 0.40 $\pm$ 0.03 \\
    \hline
    \end{tabular}
\end{table*}

To retrieve more information from these data, we model the tip-nanotube contact area $A$ employing the Hertz model in the configuration of a sphere pressed in contact with a cylinder \cite{22}. Then the friction force between the AFM tip and the BN-NT can be expressed by the following equation:

\begin{equation} 
F_{F} = \sigma \cdot A = \sigma \cdot \gamma \cdot (F_{N} + F_{Adh})^{2/3}
\end{equation}

where $\sigma$ and $F_{Adh}$ are respectively the shear strength and adhesion force between the tip and the BN-NT \cite{23}. The quantity $\gamma$ is a function of $R_{NT}$, $R_{tip}$ and the elasticity of both surfaces. The derivation of equation (1) can be found in the literature \cite{24,25}. Next, we fit the friction data obtained for different BN-NTs with equation (1) to estimate the transverse and longitudinal shear strengths $\sigma_{T}$, $\sigma_{L}$ and $F_{Adh}$. The fitted shear strengths $\sigma_{T}$ and $\sigma_{L}$ are reported in figure 2 (a) and (b) as a function of $R_{NT}$, together with the values of $\sigma_{T}$ and $\sigma_{L}$ found for CVD C-NTs and AD C-NTs, as reported in a previous work \cite{12}. The transverse shear strength of BN-NTs displays values between $0.06 \pm 0.001$ and $0.23 \pm 0.06$ GPa and is larger than $\sigma_{L}$, which varies between $0.02 \pm 0.001$ and $0.1 \pm 0.01$ GPa. The friction anisotropy, defined as $a \equiv \sigma_{T} / \sigma_{L}$, is between 1.6 and 8.4 for the studied BN-NTs. These results indicate that the measured shear strengths of CVD BN-NTs for both T and L sliding have intermediate values between those of AD C-NTs and CVD C-NTs, probably owing to a different amount of structural defects present in the nanotubes \cite{13}.

To better understand these results and the origin of the different tribological behaviour in BN-NTs as compared to C-NTs, we apply a model previously developed for C-NTs \cite{12}. For an ideal, armchair, and defect-free nanotube deposited on a flat surface, the friction anisotropy originates from the effect of a transverse swaying also called "hindered rolling" when an AFM tip slides on the nanotube perpendicularly to its axis. These transversal deformations provide a new channel for energy dissipation and increase the overall friction force between the tip and the nanotube. On the other hand, during the longitudinal sliding, such energy dissipation mechanism is absent for a perfect armchair nanotube, leading to smaller longitudinal friction forces which are mainly resulting from the "intrinsic" force needed for sliding the hard tip-nanotube contact. This anisotropy in the measured friction force is clearly demonstrated by the data reported in figure 2(a) and 2(b). However, for nanotubes with different chirality, structural defects and chemical functionalization, this additional structural inhomogeneity can couple the transverse and longitudinal motion, giving rise to larger longitudinal friction forces and smaller transverse friction forces \cite{12}. This reasoning can be formalized with an analytical model consisting of two equations \cite{12}:

\begin{equation} 
\sigma_{L} = \sigma^{int} + \alpha \cdot \sigma^{HR}
\end{equation}

\begin{equation} 
\sigma_{T} = \sigma^{int} + (1 - \alpha) \cdot \sigma^{HR}
\end{equation}

where $\sigma^{int}$ is defined as the intrinsic shear strength between a silicon AFM tip and a flat $h$-BN film (or graphite in the case of C-NT), $\sigma^{HR}$ is the shear strength resulting from the effect of "hindered rolling" and $\alpha$ is a coupling parameter between the transversal and longitudinal motion, due to the presence of structural defects, chirality or other  properties.  In a previous study it was shown that even for very different types of nanotubes the friction anisotropy as a function of $\alpha$ was approximately always following the same analytical curve \cite{12}. Thus, here we will be able to calculate $\alpha$ for each BN-NT with a given friction anisotropy. Afterwards, we apply the system of equations (2) and (3) to obtain the values of $\sigma^{int}$ and $\sigma^{HR}$ for each data presented in figure 2 (a) and (b). The average $\sigma^{int}$ for BN-NTs is found to be $0.007 \pm 0.002$ GPa and $\sigma^{HR}$ results to be $0.18 \pm 0.07$ GPa. This value of intrinsic shear strength is larger than the value of $\sim 0.005$ GPa obtained for defects-free AD-CNTs \cite{12} but smaller than the value of $0.022$ GPa obtained for CVD CNTs, which have abundant structural defects \cite{12}. This is consistent with the low amount of defects in BN-NTs and the fact that the friction coefficient of $h$-BN is larger than the value of highly ordered pyrolytic graphite (HOPG) at the room temperature \cite{26}. In Table 1, we summarize the key parameters describing the frictional properties of the investigated BN-NTs, and their sister counterpart, C-NTs \cite{12}. The friction anisotropy of BN-NTs is found to be in the range between 1.6 and 8.4, while for AD and CVD CNTs, the range is 2.5 to 13.7 and 1.5 to 5.7, respectively. These results reflect the values of transverse and longitudinal shear strengths reported in figure 2 and can be explained by either the higher friction coefficient of h-BN compared to HOPG, or an amount of defects in CVD BN-NTs smaller than in CVD C-NTs. From TEM and AFM images, we indeed find that the CVD BN-NTs are very straight, with a structural order similar to that one of AD C-NTs \cite{12}, whereas CVD C-NTs are rich of defects and usually curly. The average coupling parameter of BN-NTs, $\alpha = 0.25 \pm 0.03$, is similar to that one of CVD C-NTs but larger than in AD CNTs. Since $\alpha$ is a function of structure and surface properties, this suggests the CVD BN-NTs might have more structural defects than AD C-NTs, which can contribute to the T-L coupling. The maximum friction anisotropy for CVD BN-NTs is found to be equal to $8.4$, smaller than the value of $14$ found in AD CNTs. This could be attributed to the absence of armchair BN-NTs, in fact, molecular dynamics simulations have shown that for an armchair C-NT, the friction anisotropy can be as large as $20$ but will be only $2$ if the external wall of the C-NT is chiral \cite{13}. Finally, the average "hindered rolling" shear strength of the studied BN-NTs, $\sigma^{HR} = 0.18 \pm 0.02$ GPa, is also in between $\sigma^{HR}$ of AD and CVD C-NTs, but more similar to the value of AD C-NTs. However, we underline that the phenomenon of hindered rolling is complex and depends on many variables, such as the size of the nanotube, thickness of the tube, structural defects and adhesion forces between the AFM tip, nanotube and the substrate. Further investigations and simulations are necessary to identify the role of each parameter in determining the frictional behaviour of nanotubes on the surface.

Besides structural defects, chirality and intrinsic friction, also adhesion forces may play a role in the friction properties of BN-NTs and C-NTs. The averaged values of $F_{Adh}$ between the AFM-tip and CVD BN-NTs, AD C-NTs and CVD-CNTs are $4.8 \pm 1.3$ nN, $15.0 \pm 2.4$ nN and $5.4 \pm 0.7$ nN, respectively; however $F_{Adh}$ measured between the AFM tip and the silicon surface where the respective nanotubes were deposited are $2.2 \pm 0.3$ nN, $17.5 \pm 2.5$ nN and $8.7 \pm 1.1$ nN, respectively. Thus, if we consider the ratio $F_{Adh-Nanotube}/F_{Adh-Si}$, we obtain $2.2 \pm 1.3$, $0.86 \pm 3.5$ and $0.62 \pm 1.3$ for CVD BN-NT, AD C-NT and CVD C-NTs, respectively, showing that the tip-nanotube adhesion is the same within the error bars for these three types of nanotubes and it does not have a clear relationship with the friction anisotropy, as shown in Figure 2(c). The difference in $F_{Adh}$ on silicon can probably be ascribed to different relative humidity values during these experiments. Note that these adhesion forces are measured during the friction measurements. On the other hand, the adhesion force between a nanotube and the silicon substrate cannot be measured directly, but it can be estimated from the contact area between the nanotubes and the substrate. These nanotube-substrate adhesion forces, together with the chirality and the structural defects, might be important in determining the aforementioned friction anisotropy. Longer and larger nanotubes, with a larger contact area with the silicon surface, might be anchored more strongly on the surface and might be less prone to wobble during the AFM tip sliding. This situation, up to a certain value of nanotube-Si adhesion, is likely to affect mainly the longitudinal friction, because the adhesion will reduce the coupling between the transversal and longitudinal motions, giving rise to a smaller longitudinal shear strength and, as a consequence, larger friction anisotropy. For very large nanotube-Si adhesion forces and thick tubes, on the other hand, it is possible that the adhesion will prevent any transversal deformation, reducing the friction anisotropy coefficient to $1$. The investigated nanotubes have various lengths, $L_{NT}$, and radii, $R_{NT}$. Thus the nanotube-Si contact area, proportional to the adhesion force, will be different depending on $L_{NT}$ and $R_{NT}$, leading to different degree of anchoring of the nanotube on the substrate. For a cylinder with radius $R$ and length $L$ lying on a surface with its principal axis parallel to the substrate surface, the contact area is rectangular and can be calculated by the Hertz theory \cite{27}. This rectangular contact has a side length $L$ and a width $2b$, where $b = ( 2 F_{N} R / \pi L E^{*} )^{1/2}$, with $F_{N}$ denoting the normal load that presses the cylinder onto the substrate and $E^{*} = ( ( 1 - \nu_{1}^{2} ) / E1 ) + ( ( 1 - \nu_{2}^{2} ) / E2 ) )^{-1}$ is the effective modulus. The parameter $\nu_{1,2}$ and $E_{1,2}$ are respectively the Poisson's ratio and the Young's modulus of both objects. The normal loads $F_{N}$ ($< 4$ nN) are similar in all measurements. The radial modulus of BN-NTs and C-NTs are also similar within the range of RNT under investigation \cite{23,28}. Therefore the contact area, $L \cdot 2b$, is approximately proportional to $L \cdot ( R / L )^{1/2} \approx ( L \cdot R )^{1/2}$ and should be roughly proportional to the adhesion force between the cylinder and the substrate. In figure 3(a), we plot the friction anisotropy of all BN-NTs and C-NTs as a function of $( L_{NT} \cdot R_{NT} )^{1/2}$, whereas figure 3(b) shows a zoom-in of the friction anisotropy for only BN-NTs. In figure 3(a) and 3(b), a quasi-linear dependence between the friction anisotropy (open symbols) and $( L_{NT} \cdot R_{NT} )^{1/2}$ demonstrates that a larger adhesion force between the nanotube and the substrate indeed reduces the T-L coupling and increases the friction anisotropy. Figures 3(c) and 3(d) show the values of $\sigma_{T}$ and $\sigma_{L}$ as a function of $( L_{NT} \cdot R_{NT} )^{1/2}$, respectively. These data indicate that while $\sigma_{T}$ is approximately constant for different contact areas, i.e., for different nanotube-Si adhesion forces, $\sigma_{L}$ decreases quite strongly with increasing $( L_{NT} \cdot R_{NT} )^{1/2}$. This result supports the hypothesis that the main effect of a larger adhesion is the reduction of transversal deformations during longitudinal tip sliding. Finally, we remark that the solid symbols in figure 3, which represent the AD C-NTs with extremely large friction anisotropy ($a > 12$), have been identified in a previous work as armchair or very symmetric C-NTs.

\section*{Conclusion}

In summary, by means of atomic force microscopy we have investigated the frictional properties of individual multi-wall CVD BN-NTs deposited on a silicon substrate, and compared their behaviour with previous measurements performed on C-NTs grown by the AD and CVD methods. We find that the transverse and longitudinal shear strengths of CVD BN-NTs are larger than those of AD C-NTs but smaller than in CVD C-NTs. These findings are related to a smaller amount of defects in CVD BN-NTs than in CVD C-NTs, and to the larger friction coefficient for $h$-BN than for HOPG. The friction anisotropy of the studied BN-NTs is found to be between $1.6$ and $8.4$. Finally, using Hertz's contact mechanics we calculate the nanotube-Si contact area, which is proportional to $( L_{NT} \cdot R_{NT} )^{1/2}$ and to the nanotube-Si adhesion, and find that the friction anisotropy increases quasi linearly with $( L_{NT} \cdot R_{NT} )^{1/2}$. This result is explained by considering that a stronger adhesion between the nanotube and the substrate likely reduces the transversal deformations during longitudinal tip sliding leading to smaller longitudinal shear strengths and hence to larger friction anisotropy. Understanding how nano-objects interact at the nanoscale is important in developing robust and reliable nanodevices such as nano-electro-mechanical system (NEMS) and nanocomposites materials. C-NTs have been proposed as components for nano-devices, nanotube bearings for wear-free surfaces \cite{29} or nanoswitches \cite{30}, and reinforcement in composite materials \cite{31}. On the other hand, BN-NTs, with similar or better tribological properties will be a more appropriate choice for applications where high temperatures are required. Our results provide a better and fundamental understanding of the frictional properties of BN-NTs at the nanoscale.

\section*{Acknowledgements}

H.-C. C. and E. R. acknowledge the financial support of the Office of Basic Energy Sciences DOE (DE-FG02-06ER46293). E. R. acknowledges the National Science Foundation NSF (DMR-0120967 and DMR-0706031) for partial support.


\end{document}